\documentclass[iop,apj]{emulateapj}

\def\lea{\mathrel{<\kern-1.0em\lower0.9ex\hbox{$\sim$}}}
\def\gea{\mathrel{>\kern-1.0em\lower0.9ex\hbox{$\sim$}}}
\newcommand{\lta}{{\>\rlap{\raise2pt\hbox{$<$}}\lower3pt\hbox{$\sim$}\>}}
\newcommand{\gta}{{\>\rlap{\raise2pt\hbox{$>$}}\lower3pt\hbox{$\sim$}\>}}

\begin{document}
\title{The Connection Between X-ray Binaries and Star Clusters in NGC 4449}

\author{Blagoy Rangelov\altaffilmark{1}, Andrea H. Prestwich\altaffilmark{2}, and Rupali Chandar\altaffilmark{1}}
\altaffiltext{1}{Department of Physics \& Astronomy, The University of Toledo, 2801 West Bancroft Street, Toledo, OH 43606}
\altaffiltext{2}{Harvard-Smithsonian Center for Astrophysics, 60 Garden Street, Cambridge, MA 02138}
\email{blagoy.rangelov@gmail.com}

\slugcomment{The Astrophysical Journal, in press}
\shorttitle{XRBs and Star Clusters in NGC 4449}
\shortauthors{Rangelov et al. 2011}

\begin{abstract}

We present 23 candidate X-ray binaries with luminosities down to
$1.8\times10^{36}$ erg s$^{-1}$, in the nearby starburst galaxy NGC
4449, from observations totaling 105 ksec taken with the ACIS-S
instrument on the \emph{Chandra Space Telescope}. We determine count rates,
luminosities, and colors for each source, and perform spectral fits
for sources with sufficient counts. We also compile a new catalog of
129 compact star clusters in NGC 4449 from high resolution, multi-band
optical images
taken with the \emph{Hubble Space Telescope}, doubling the number
of clusters known in this galaxy.  The $UBVI$,$H\alpha$ luminosities of
each cluster are compared with predictions from stellar evolution
models to estimate their ages and masses. 
We find strong evidence for a population of very young
massive, black-hole binaries, which comprise nearly 50\% of
the detected X-ray binaries in NGC~4449. 
Approximately a third of these remain within their
parent star clusters, which formed
$\tau \lea 6-8$~Myr ago, while others have 
likely been ejected from their parent clusters.
We also find evidence for a population of 
somewhat older X-ray binaries, including
both supergiant and Be-binaries, which
appear to be associated with somewhat older
$\tau \approx 100-400$~Myr star clusters,
and one X-ray binary in an ancient ($\tau \approx 10$~Gyr) 
globular cluster.
Our results suggest that detailed information
on star clusters can significantly improve
constraints on X-ray binary populations in star-forming galaxies.

\end{abstract}


\section{Introduction}

Images of nearby starburst galaxies taken in X-rays with the
\emph{Chandra Space Telescope} are spectacular, showing a
multitude of bright point sources. It is now generally accepted that
most of these sources are high mass X-ray binaries (HMXBs) produced during
recent star formation.
HMXBs are binaries where one member of the system is a compact object,
either a black hole or neutron star, and the other is a young, massive
star. X-ray emission is produced as material is accreted from the young
``donor'' star onto the compact object. 
HMXBs can be divided into two general categories 
based on the type of donor star: 1) a Be star (Be/X-ray binary), or
2) a supergiant star (SG/X-ray binary).

Previous studies have suggested an intriguing connection between
HMXBs and young stellar clusters: many
of the former are found close to, but not coincident with, young star
clusters.
These observations
are broadly consistent with a scenario
where X-ray binaries (XRBs) form in star clusters,
but have sufficiently large velocities that they are expelled from
their parent cluster (\citet{Zezas2002}, \citet {Kaaret2004}). 
There are three mechanisms that might displace HMXBs from
their birthsites.
The binary could be given a ``kick'' during an
asymmetric supernova explosion that forms the black hole
or neutron star, or it could be
ejected via dynamical interactions with other stars in a dense cluster core
\citep{McSwain2007}.
A third scenario that could explain the observed spatial
displacement between HMXBs and young star clusters is
that X-ray binaries formed in clusters that have since dispersed.

In this paper, we investigate the relationship between HMXBs
and star clusters using the best X-ray and optical
observations currently available for the starburst
galaxy NGC 4449. 
We analyse observations taken with the 
\emph{Chandra Space Telescope} to select X-ray  binaries
in NGC~4449, and optical images taken with the 
\emph{Hubble Space Telescope} 
($HST$)
to detect star clusters and to estimate
their ages and masses.
An optical color image of NGC~4449 based on the
$HST$ images is shown in Figure~1.
Our primary goals are to constrain the nature of X-ray binaries
and in particular HMXBs, and to explore the relationship
between HMXBs and star clusters in NGC~4449.

The rest of this paper is organized as follows. In Section 2 we
present the X-ray observations from \emph{Chandra}, including data
reduction, source detection and some basic models used for
interpretation.
In Section 3 we present the optical observations
from the $HST$ and a new
catalog of compact star clusters, including their
integrated colors and luminosities, and we derive their
masses and ages.
The mass and age distributions of the clusters and what
they tell us about the formation and disruption of the
clusters are the subject of Section~4, while
Section~5 presents 
the spatial correlation between the
clusters and the candidate HMXBs.
Section~6 synthesizes these results to constrain
the nature of the HMXBs in NGC~4449, and Section~7
summarizes the main results of this work.

\begin{figure*}[htp]
\epsscale{1}
\plotone{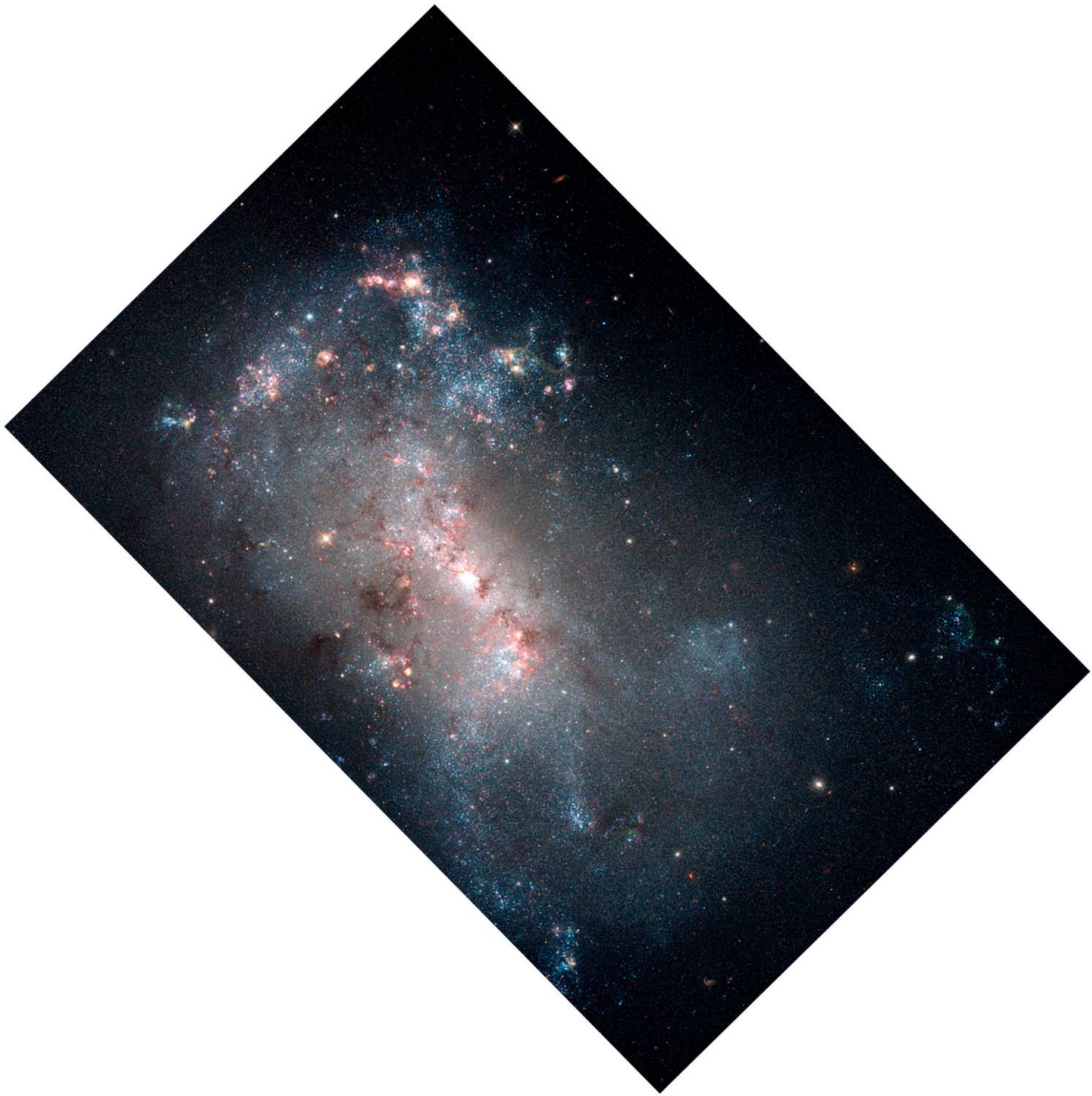}
\caption{Color image of NGC 4449 produced using the $HST$/ACS observations described
in Section~3.  The F435W ($\approx B$ band) image is shown in blue, 
the F555W ($\approx V$ band) image in green, and the red image combines the
F814W ($\approx I$ band) and F658N (H$_{\alpha}$) fiters.
North is up and east is to the left. This image is $5\arcmin$ 
along the long side. (Credit: NASA, ESA, A. Aloisi (STScI/ESA), and The Hubble Heritage (STScI/AURA)-ESA/Hubble Collaboration)
\label{fig-NGC4449}}
\end{figure*}

\section{X-ray Observations from \emph{Chandra}}

NGC~4449 is an irregular, star-forming galaxy located
at a distance of $3.82 \pm 0.27$~Mpc \citep{Annibali2008}.
With an integrated magnitude of $M_B=-18.2$ it is somewhat more lumninous
than the Large Magellanic Cloud \citep{Hunter1997}.
NGC 4449 has a current star formation rate of $\approx 1.5~M_{\odot}$~yr$^{-1}$
\citep{Thronson1987} and a near-solar present-day gas
abundance ($12+\mbox{log}[O/H]=8.83$; \citet{GS1998}).

\subsection{Data and Reduction}

We use three sets of archival $Chandra$ observations of NGC~4449, with
integration times of 30~ksec (ObsID: 2031, PI: Heckman), 
15~ksec (ObsID: 10125. PI: Long),
and 60~ksec (ObsID: 10875, PI: Long), to detect X-ray point sources
in NGC~4449. Basic information for the three sets of observations 
is given in Table~1. 
The data were taken with the Advanced CCD
Imaging Spectrometer (ACIS) instrument on the \emph{Chandra} telescope
on February 4, 2001 in ``faint'' mode (ObsID: 2031), and March 4 and
7, 2010 in ``vfaint'' mode (ObsIDs: 10125 and 10875).  The galaxy
was positioned on the back-illuminated S3 CCD chip. 
We processed the data using the \emph{Chandra} Interactive Analysis of
Observations (CIAO) software (version 4.2) and \emph{Chandra}
Calibration Data Base (CALDB) version
4.3.0\footnote{http://cxc.harvard.edu/ciao/}, and  restricted
the data to the energy range between 0.3-8 keV.
The observations were filtered in three
energy bands -- 0.3-1 keV (soft), 1-2 keV (medium) and 2-8 keV (hard).

\subsection{Source Detection and X-ray Properties}

We use CIAO's Mexican-hat wavelet source detection routine
\emph{wavdetect} \citep {Freeman2002} to create source lists. Wavelet
scales of 1.4, 2, 4, 8, and 16 pixels and a detection threshold of
10$^{-6}$ were used, which typically results in 1 spurious detection
per million pixels. The output sources were examined visually to
verify each detection, and to correct the source catalog when multiple
detections occurred. A catalog of X-ray point sources detected in the
{\em Chandra} observations is presented in Table~\ref{tbl-X-ray},
and the locations of these sources are shown in Figure~\ref{fig-NGC4449}.

\begin{figure*}
\epsscale{1}
\plotone{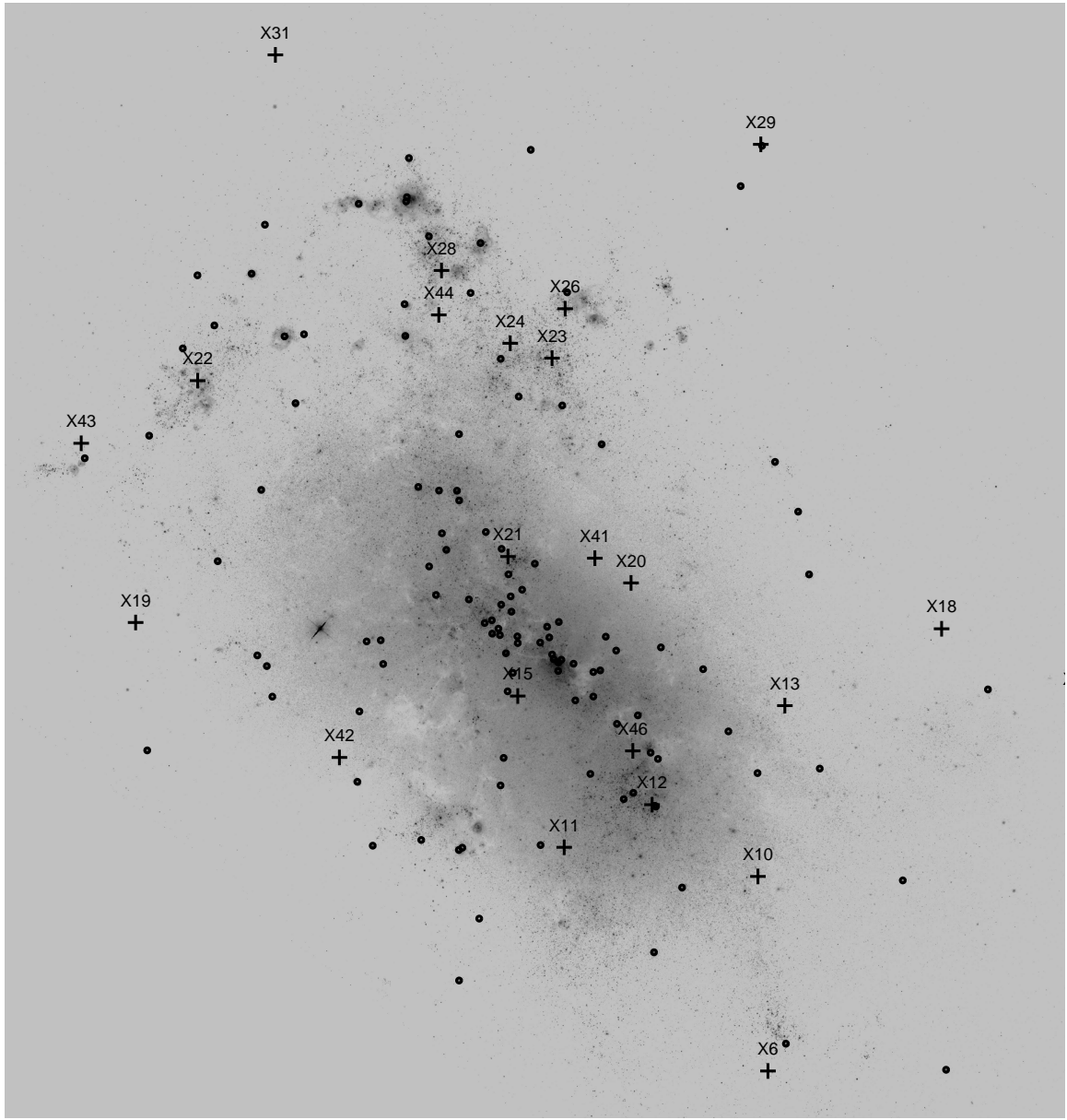}
\caption{$V$ band (F555W filter) image of NGC 4449 taken with the ACS camera on the $HST$. 
North is up and east is to the left. The locations of star clusters selected from optical $HST$ images
are shown as circles. The locations of X-ray sources selected from $Chandra$ X-ray observations are labeled and shown as crosses.
\label{fig-NGC4449}}
\end{figure*}

A number of measurements and estimates are made for each source $-$ total number of counts and the
counts measured in the soft, medium, and hard energy bands. 
We calculate two X-ray colors, a
``soft'' color defined as $H1 = (M - S)/T$, and a ``hard'' color defined as
$H2 = (H - M)/T$, where $S$, $M$ and $H$ are the total counts measured
in the soft, medium and hard bands \citep{Prestwich2003}, and $T$ is
the total number of counts in all three bands. 
The luminosity of each source is estimated by fitting its
spectrum with a power law model with a photon index $\Gamma = $ 1.5. 
These values are compiled in Table~\ref{tbl-X-ray}. 

\begin{deluxetable}{rccc}
\tablecaption{$Chandra$ observations\label{tbl-Chandra}}
\tablewidth{0pt}
\tablehead{
\colhead{ObsId} & \colhead{Date} & \colhead{PI} & \colhead{Exposure}\\
\colhead{} & \colhead{} & \colhead{} & \colhead{(ks)}
}
\startdata
2031 & 2001-02-04 & Heckman & 30 \\
10125 & 2009-03-04 & Long & 15 \\
10875 & 2009-03-07 & Long & 60 \\
\enddata
\end{deluxetable}

There are eleven sources which have sufficient counts ($>$ 50) for a
crude spectral fit. We were able to fit ten of these with simple X-ray
spectral models, as summarized in Table \ref{tbl-models}. One source, X15,
is in a region with high background from diffuse emission and we were
unable to obtain a satisfactory fit. X-ray spectra and responses
(including sensitivity of the instrument and CCD) were extracted using
standard CIAO software. Spectra were grouped for a minimum of 15
counts per bin.  The fits were performed using XSPEC V 12.0 over an
energy range 0.2$-$8.0 keV. For each source, we tried three models: a
simple power law (PL), a MEKAL model\footnote{An emission spectrum from hot
diffuse gas, e.g. \citet{Liedahl1995}}, and a multi-color disk
(MCD\footnote{A superposition of multi-temperature blackbody spectra
expected from optically-thick accretion disk; the model only
constrains the temperature of the inner disk $T_{in}$,
e.g. \citet{Mitsuda1984}}) model. A fixed foreground Galactic column
$n_{H}=1.5\times10^{20}$ cm$^{-2}$ was assumed in each fit. The fits
allow for an additional variable column due to absorption intrinsic
to the source.
The best fit model and parameters for each source are given in Table~\ref{tbl-models}.
Observed X-ray fluxes and estimated luminosities  were calculated for the 0.3 $-$ 8 keV energy range (both uncorrected for absorption) using the best fit parameters in the table. 

\begin{deluxetable}{lccrrc}
\tabletypesize{\scriptsize}
\tablecaption{X-ray source catalogue\label{tbl-X-ray}}
\tablewidth{0pt}
\tablehead{
\colhead{ID} & \colhead{RA} & \colhead{DEC} & \colhead{Soft} & \colhead{Hard}  & \colhead{L$_{X}$\tablenotemark{1}}\\
\colhead{} & \colhead{} & \colhead{} & \colhead{Color} & \colhead{Color} & \colhead{(erg s$^{-1}$)}
}
\startdata
X8   & 187.00308 & 44.07569 & 0.20 & 0.00  & 4.55E+36 \\
X14 & 187.00572 & 44.09145 & 0.15 & -0.50  & 9.41E+36 \\
X18 & 187.01634 & 44.09551 & 0.01 & -0.09  & 7.68E+37 \\
X13 & 187.02855 & 44.09120 & 0.31 & -0.42  & 9.21E+36 \\
X6   & 187.02988 & 44.07074 & 0.34 & 0.14  & 1.47E+37 \\
X29 & 187.03043 & 44.12264 & 0.04 & -0.07  & 3.34E+37 \\
X10 & 187.03065 & 44.08164 & -0.33 & -0.16  & 5.67E+37 \\
X12 & 187.03891 & 44.08566 & -0.06 & -0.32  & 3.01E+38 \\
X20 & 187.04055 & 44.09808 & 0.23 & -0.13  & 1.01E+38 \\
X11 & 187.04576 & 44.08327 & 0.34 & -0.26  & 3.05E+36 \\
X23 & 187.04671 & 44.11066 & -0.24 & -0.26  & 2.25E+37 \\
X15 & 187.04985 & 44.09188 & -0.44 & -0.04  & 7.66E+37 \\
X24 & 187.04997 & 44.11149 & 0.02 & -0.29  & 5.81E+37 \\
X21 & 187.05013 & 44.09955 & -0.28 & -0.25  & 1.19E+37 \\
X28 & 187.05533 & 44.11557 & -0.01 & -0.32  & 6.93E+37 \\
X31 & 187.06829 & 44.12765 & -0.08 & 0.41  & 9.53E+36 \\
X22 & 187.07433 & 44.10939 & 0.41 & -0.13  & 8.59E+38 \\
X19 & 187.07917 & 44.09585 & -0.90 & -0.02  & 2.01E+37 \\
X41 & 187.04336 & 44.09946 & -1.00 & 0.00  & 1.43E+37 \\
X42 & 187.06327 & 44.08829 & -0.41 & -0.57  & 1.78E+36 \\
X43 & 187.08341 & 44.10589 & 0.41 & 0.09  & 3.36E+36 \\
X44 & 187.05553 & 44.11308 & 0.13 & 0.09  & 1.13E+36 \\
X46 & 187.04041 & 44.08867 & -0.20 & -0.23  & 7.59E+36 \\
\hline
X26\tablenotemark{2} & 187.04569 & 44.11343 & 0.11& -0.35 & 2.57E+38 \\
\enddata
\tablenotetext{1}{X-ray luminosities (0.3 $-$ 8 keV range) are derived by fitting a power law 
($\Gamma$ = 1.5) to the data. An assumed Galactic $n_H = 1.5\times10^{20}$ cm$^{-2}$ is applied.}
\tablenotetext{2}{Supernova remnant.}
\end{deluxetable}

\begin{deluxetable*}{lccccccccc}
\tabletypesize{\scriptsize}
\tablecaption{Model parameters for X-ray sources with $>$ 50 counts\label{tbl-models}}
\tablewidth{0pt}
\tablehead{
\colhead{ID} &  \colhead{Net} & \colhead{Best Fit} & \colhead{$n_H$} & \colhead{Photon} &  \colhead{T$_{in}$\tablenotemark{1}} & \colhead{T\tablenotemark{2}} & \colhead{$\chi^{2}$ / d.o.f.} & \colhead{Flux\tablenotemark{3}} & \colhead{L$_{X}$\tablenotemark{3}} \\
\colhead{} &  \colhead{Counts} & \colhead{Model} & \colhead{($\times10^{22}$ cm$^{-2}$)} & \colhead{index} &  \colhead{(keV)} & \colhead{(keV)} & \colhead{} & \colhead{(erg s$^{-1}$ cm$^{-2}$)} & \colhead{(erg s$^{-1}$)}
}
\startdata
X18 & 141.3 & ABS*PL & (4.5$^{+8.0}_{-4.0}$)$\times10^{-2}$ & 1.5$\pm 0.3$&  $-$ &  $-$ & $~$2.4 / 7 & 3.65E-14 &6.30E+37  \\
X29 & 84.8 & ABS*PL & (2.2$^{+12.5}_{-2.2}$)$\times 10^{-2}$& 1.26$^{+0.6}_{-0.4}$ & $-$ & $-$ & $~$0.43 / 3 & 2.48E-14 & 4.30E+37  \\
X10\tablenotemark{a} & 172.7 & ABS*(PL+MEKAL) & (9.5$^{+40}_{-9.0}$)$\times 10^{-3}$& 2.1$\pm0.4$ & $-$ & 0.46$^{+0.16}_{-0.4}$ & 7.0 / 7  & 2.72E-14 & 4.70E+37 \\
X12 & 1197.7&  ABS*MCD & (4.0$^{+4.0}_{-8.5}$)$\times 10^{-3}$& $-$ & 0.60$\pm 0.5$ & $-$ & $~$85.9 / 71 & 1.58E-13 & 2.76E+38 \\
X20 & 201.4 &  ABS*PL & 0.19$\pm0.1$ & 1.7$^{+0.4}_{-0.3}$ & $-$ & $-$ & 13.72 / 11 & 4.80E-14 & 8.35E+37 \\
X23\tablenotemark{b} & 106.0 & ABS*MCD &  0.037$^{+0.2}_{-0.04}$  & $-$ & 0.37$\pm 0.13$ & $-$ & $~$5.28 / 4 & 1.13E-14 & 1.95E+37 \\
X24 &186.8  &  ABS*PL & 0.19$^{+0.06}_{-0.08}$ & 2.4$^{+0.4}_{-0.3}$  & $-$ & $-$ & $~$7.54 / 10 & 3.20E-14 & 5.66E+37 \\
X21 & 58.9 & ABS*MCD & 0.21$^{+0.38}_{-0.38}$ & $-$ & 0.23$\pm 0.1$ & $-$ & $~$2.89 / 2$~$ & 6.00E-15 & 1.68E+37 \\
X28\tablenotemark{a} & 309.8 & ABS*MCD & 0.04$^{+0.04}_{-0.03}$ & $-$ & 0.64$\pm 0.1$ & $-$ & 24.7 / 17 & 3.98E-15 & 6.98E+37 \\
X22 & 1411.6& ABS*PL & 0.6$\pm 0.07$ & 1.9$^{+0.16}_{-0.14}$& $-$ & $-$ & 86.73 /  85 & 4.34E-13 & 7.62E+38 \\
\enddata
\tablenotetext{1}{The inner temperature of the multi-color disk model.}
\tablenotetext{2}{MEKAL model temperature.}
\tablenotetext{3}{Observed X-ray fluxes and estimated luminosities for the 0.3 $-$ 8 keV energy range, uncorrected for absorption.}
\tablenotetext{a}{Possible line emission.}
\tablenotetext{b}{Excess flux just over 1 keV.}
\end{deluxetable*}

\subsection{X-ray Colors and Models}

After examining the resulting X-ray source catalog, we eliminate some
sources from further consideration in this work for the following
reasons. 
First, we remove all sources located beyond the optically
luminous portion of NGC 4449 (R $\thicksim$ 3-4 kpc), since these are
almost certainly background galaxies and not associated with NGC 4449
itself. 
We also eliminate one source that is a confirmed supernova
remnant in NGC 4449 \citep{PF2003}, and another which appears to be a spurious
detection of diffuse emission near the center of the galaxy. Two
additional X-ray sources
are coincident with foreground stars and were
also eliminated. Our final catalog contains 23
sources that we believe are X-ray binaries.
These are listed in Table \ref{tbl-X-ray}.

\begin{figure*}
\epsscale{1}
\plotone{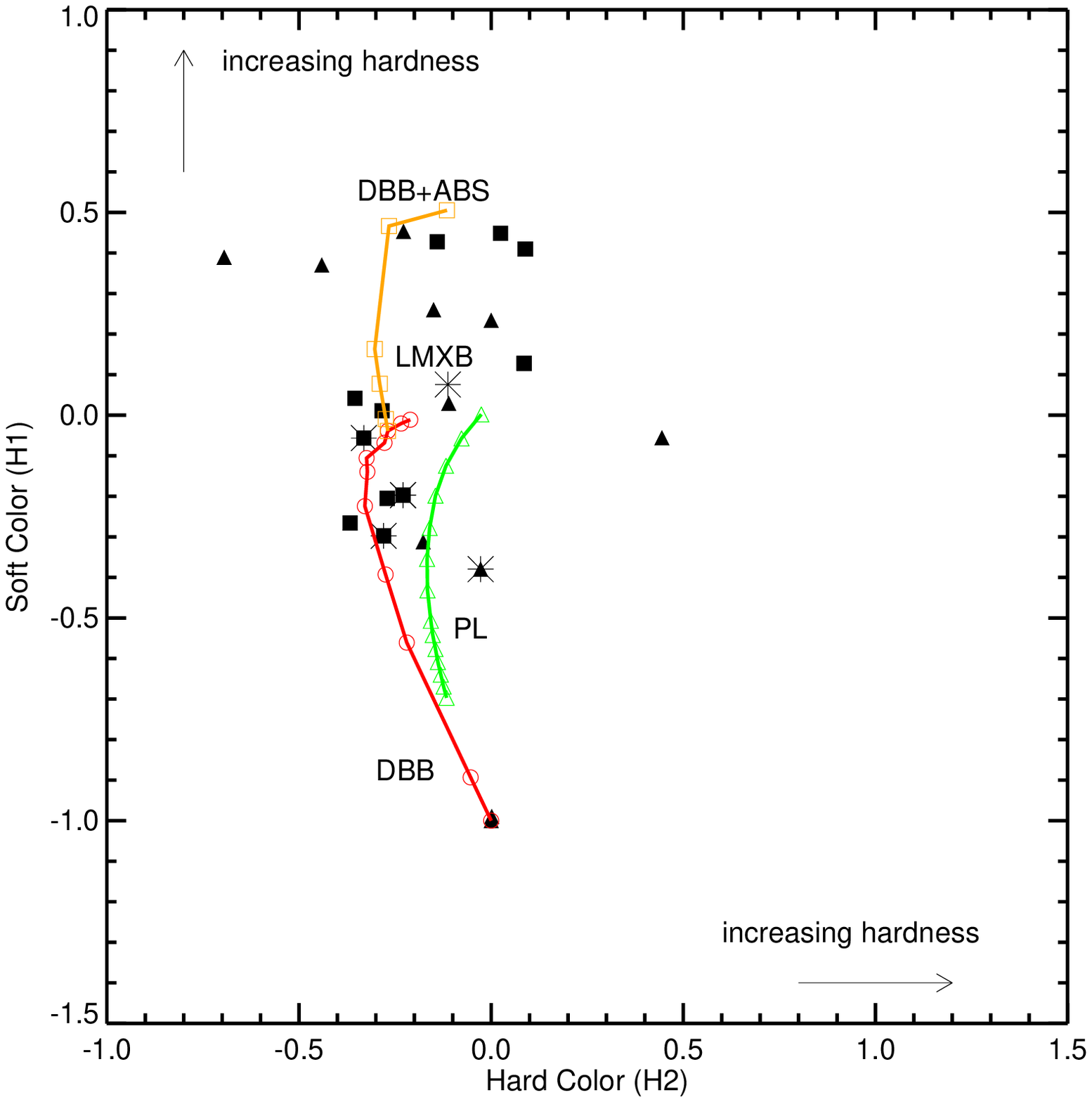}
\caption{X-ray color-color diagram of X-ray binaries in NGC~4449.
The H1 and H2 colors are defined in Section~2.
Candidate high mass black hole binaries are shown as squares,
and the rest of the sample is shown as triangles.
The asterisks show the X-ray sources that are coincident with clusters.
Theoretical tracks are shown as the open symbols connected with solid lines. The green triangles represent a power law (PL) with an increasing photon index from 1.0 to 3.0, and the red circles show predictions for diskblack-body models (DBB) with increasing temperature from 0.1 to 1.0 keV. The orange squares are a disk black-body model with T=0.9 keV and an increasing hydrogen column density (DBB+ABS) $n_H$ in steps of 0, $1\times10^{20}$, $5\times10^{20}$, $1\times10^{21}$, $5\times10^{21}$, and $1\times10^{22}$ cm$^{-2}$.
See text for details.
\label{fig-X-ray-cc}}
\end{figure*}

The observed H1 vs. H2 colors
of the X-ray binaries are shown in Figure \ref{fig-X-ray-cc}. Two
models are shown for reference. The red line shows 
predictions from disk blackbody
models with temperatures ranging from $0.1-1.0$ keV. Black hole binaries
typically have disk blackbody temperatures of kT $\thicksim 1.0$ keV. The
orange line shows the effect of adding absorption to the disk
blackbody models. The green line represents a power law with increasing
photon index from $1.0-3.0$. 
Accreting low mass neutron star binaries
typically have absorbed power law spectra with photon index $\thicksim
2.0$. Figure \ref{fig-X-ray-cc} suggests that NGC 4449 has a mix of X-ray binary populations
typical of a galaxy that has had on-going star formation, with some
sources better following disk black-body models and others
in the absorbed portion of the diagram.
The nature of the X-ray binaries will be discussed in more
detail in Section~6.

\section{Optical Observations from \emph{HST}}

It has been suggested that the dense environments found in compact
star clusters may be very efficient in producing XRBs (e.g., \citet{McSwain2007}).
Regardless of whether or not XRBs form within compact clusters, the
clusters are good tracers of star formation in galaxies,
and hence can provide important constraints on the XRB population.
The goals of this section are two-fold: 
(1) select a new catalog of compact star clusters in NGC 4449, to
correlate with the locations of the XRBs, and 
(2) measure luminosities and colors
for any point source coincident with an XRB and hence
likely to be the donor star (discussed in Sections~5.2 and 6.4).
While a catalog of clusters in NGC 4449
already exists \citep{GHG2001}, it is based on partial imaging taken
with the WFPC2 instrument on board $HST$, and there are now deeper,
higher resolution data with full coverage of NGC~4449 available from
the Wide Field Channel (WFC) camera of the Advanced Camera for Surveys
(ACS) instrument.

\subsection{Data and Photometry}

ACS/WFC imaging of two positions within NGC 4449 was taken in the
F435W ($\approx B$), F555W ($\approx V$ band), F814W ($\approx I$
band), and the F658N ($H\alpha$) filters, on November 10-11, 2005
(Proposal ID GO: 10585, PI: Aloisi).  Four individual exposures were
taken in each filter at each pointing. The ACS/WFC has a pixel scale
of 0.$\arcsec$049, or a projected scale of 18.4 pc per arcsecond at the
assumed distance of $3.82\pm0.27$ Mpc \citep{Annibali2008} to NGC 4449.  
We downloaded the ACS data from the Hubble Legacy Archive\footnote{http://hla.stsci.edu/}
(HLA).  The HLA combines the individual flatfielded exposures for a
specific filter together using the PYRAF task $Multidrizzle$, and
outputs geometrically corrected images.
For the NGC 4449 observations, the HLA used stars from the U.S. Naval Observatory
catalogue to astrometrically correct the images.

We also use available F336W ($\approx U$) band imaging of two
pointings within NGC~4449, taken with the WFPC2 camera (Proposal ID GO: 6716, PI: Stecher). 
Each position has two exposures.  
The WFPC2 has four CCDs $-$ the Planetary
Camera (PC) has a scale of 0.$\arcsec$0456~pix$^{-1}$, and the three
Wide Field (WF) CCDs have a scale of 0.$\arcsec$0996~pix$^{-1}$.  Note
that there are two sets of images for each WFPC2 observation available
in the HLA: a combined WFPC2 image including all four CCDs with the PC
resampled to the same resolution as the three WF CCDs, and an image of
only the PC,
at its original pixel scale.  We
have used the combined WFPC2 image when our cluster candidates were in
one of the three WFs and WFPC2-PC for objects located in the PC.
Details on pointings and exposure times for the $HST$ data used here
are given in Table~\ref{tbl-HST}.

\begin{deluxetable}{ccll}
\tablecaption{$HST$ Images\label{tbl-HST}}
\tablewidth{0pt}
\tablehead{
\colhead{$HST$ Instrument} & \colhead{Proposal ID} & \colhead{Filter} & \colhead{Exposure time (s)}}
\startdata
WFPC2 & 6716 & F336W & 2$\times$520 \\
ACS/WFC & 10585 & F435W & 4$\times$3660 (PosA) \\
&&& 4$\times$3478.91 (PosB) \\
&& F555W & 4$\times$2460 \\
&& F658N & 4$\times$360 \\
&& F814W & 4$\times$2060\\
\enddata
\end{deluxetable}

We identified $\thicksim$200,000 sources in each ACS pointing (in the $V$ band),
using the IRAF\footnote{IRAF is distributed by the National Optical Astronomy 
Observatories, which are operated by the Association of Universities for 
Research in Astronomy, Inc., under cooperative agreement with the 
National Science Foundation.} DAOFIND task.
These include star clusters and bright, individual stars in NGC~4449, 
as well as
some foreground stars and background galaxies.
We perform circular aperture photometry of all detected sources using
radii of 1 and 3 pixels and background annuli of 8 and 13 pixels,
using the PHOT task in IRAF.
We use zeropoints on the VEGAMAG system taken from
Table 9 in \citet {Holtzman1995a} for the WFPC2/F336W filter, 
and from Table~10 in \citet {Sirianni2005} for the ACS observations.
Corrections for inefficiency in the charge transfer were determined
for the WFPC2/F336W photometry based on the formulae given by Dolphin (2000; 
we have used the most recent characterization available from A. Dolphin's website 
http://purcell.as.arizona.edu/wfpc2\_calib/). We also applied aperture corrections 
to extrapolate from our fixed aperture radius to the total magnitude in each filter,
based on the concentration index ($CI$, the magnitude difference
measured  between a radius of 1 and 3 pixels) measured for each source.
We obtained a linear fit between the measured $CI$ and aperture corrections for
$\approx$20 hand-selected, relatively isolated sources, 
and used this fit to estimate the 
aperture corrections for the rest of the sources.

\subsection{Color Magnitude Diagram of Donor Stars}

\begin{figure}
\epsscale{1.2}
\plotone{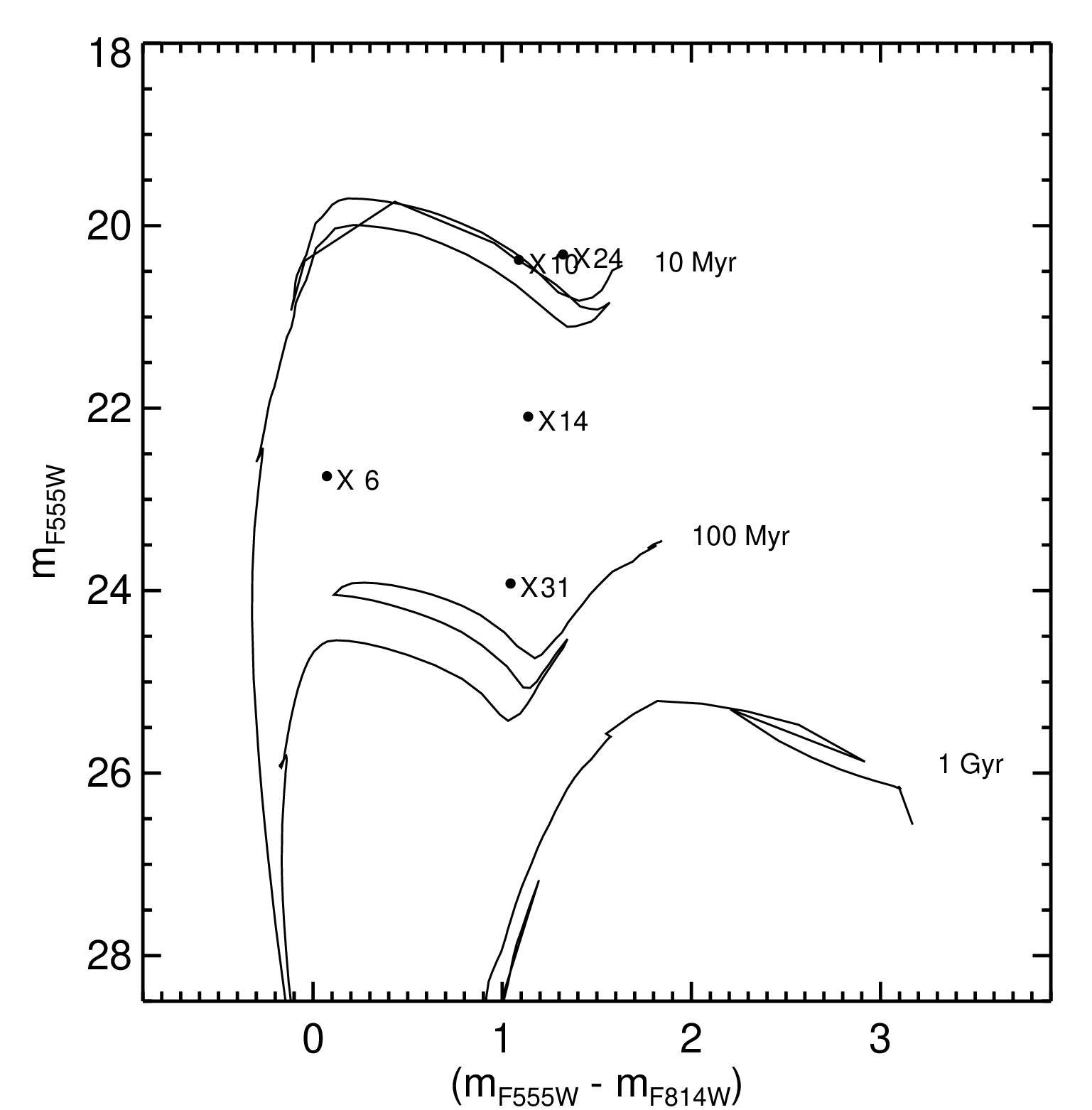}
\caption{Color magnitude diagram of optical point sources which are unique counterparts to X-ray sources.
The solid lines are Padova isochrones (for $Z = 0.02$) of 10 Myr, 100 Myr and 1 Gyr (from left to right).
\label{fig-cmd-stars}}
\end{figure}

The colors and luminosities of individual stars give an estimate of
their masses and hence a constraint on their ages. 
In several cases, the optical images reveal a single point source
within the positional uncertainty of an XRB, which is likely
the donor star.
Although a full analysis of the 
spectral energy distributions for
these stars is beyond the scope of this paper and will be presented
in a future work, we can use our photometry to broadly assess their ages.
In Figure \ref{fig-cmd-stars}, we compare the measured luminosities and colors
for these matched optical sources with solar metallicity stellar isochrones with ages
of $10$~Myr, 100~Myr, and 1~Gyr from the Padova group (\citet{Girardi2008}; \citet{Marigo2008}). 
These provide a rough guide to the ages/types of the candidate donor stars, 
and suggest that the bright stellar sources associated with X10 and X24 are supergiant stars,
while those associated with X6, X31, and X14 are significantly fainter
and hence may be somewhat older.

\subsection{Cluster Selection}

We measure the sizes of all detected objects using the 
\emph{Ishape} software \citep{Larsen1999}.
This gives a better measure than CI of the sizes of relatively bright,
isolated star clusters. \emph{Ishape} convolves analytic profiles with the PSF,
and determines the best fit to each source.
We assume a KING30 profile, a single mass \citet{King1966} model
with a fixed ratio of tidal to core radius of 30 and fit the data
within a radius of 5 pixels.
For more details on Ishape and estimating the sizes
of clusters, see \citet{Larsen1999}.

Most star clusters are slightly broader than the PSF at the distance of NGC~4449.
The biggest challenge in selecting compact 
star clusters from the entire source list
is to separate them from chance blends and superpositions in the crowded
star forming regions. 
We use the following critieria to select star cluster candidates:
(1)  $m_V \leq 22$ (M$_{V} \approx -6$ at the assumed distance of 3.82~Mpc for NGC~4449);
(2) $CI > 1.3$; (3)  FWHM $>$ 0.2 pixels, i.e. at least 0.2~pixels
broader than the PSF, as measured by Ishape; and
(4) no ``neighbors'', i.e. other object detected within a 5 pixel radius.
A final by-eye inspection was made to throw out some remaining blends
(at the $\approx20-30\%$ level).
The final catalog contains 129 candidate star clusters,
double the number published previously by \citet{GHG2001}.  Basic information
for the selected clusters, including their locations and measured photometry,
is given in Table~\ref{tbl-clusters}.
The cluster locations are shown in  Figure~\ref{fig-NGC4449}.

\begin{deluxetable}{cccrrc}
\tablecaption{Star cluster catalogue\label{tbl-clusters}}
\tablewidth{0pt}
\tablehead{
\colhead{Number} & \colhead{RA} & \colhead{DEC} & \colhead{\emph{U-B}} & \colhead{\emph{V-I}} & \colhead{Age\tablenotemark{1}} \\
\colhead{} & \colhead{} & \colhead{} & \colhead{} & \colhead{} & \colhead{log($\tau$/yr)}
}
\startdata
1 & 187.05396 & 44.083114 &  -2.106 & -0.295 & 6.40 \\
2 & 187.05370 & 44.083248 & -2.195 & -0.395 & 6.02 \\
3 & 187.06067 & 44.083354 & -0.303 & 1.080 & 9.30 \\
\nodata
\enddata
\tablecomments{This table is published in its entirety in the electronic edition of the  \emph{Astrophysical Journal}.  A portion is shown here for guidance regarding its form and content.}
\tablenotetext{1}{Typical uncertainties are approximately $\pm0.3$ in log~$\tau$.}
\end{deluxetable}

\subsection{Cluster Age, Mass and Size Estimates}

\begin{figure}
\epsscale{1.1}
\plotone{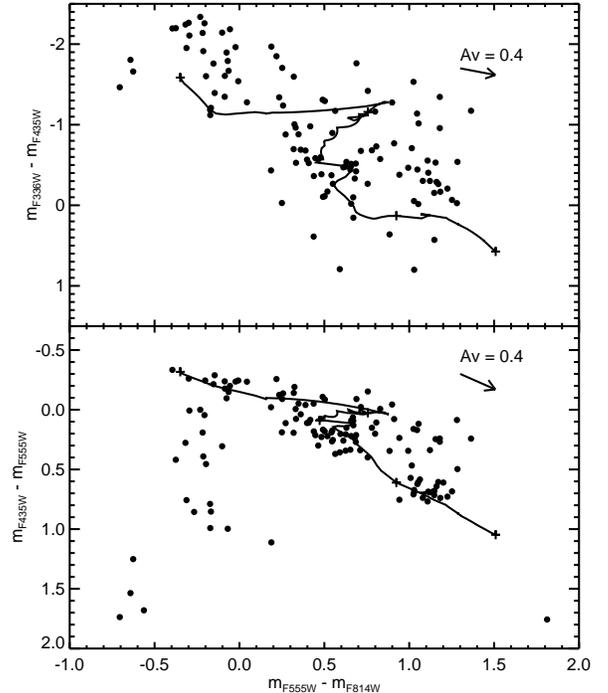}
\caption{Optical two-color diagrams of all star clusters in our sample. 
The solid line shows predictions from the single stellar population models of CB07
and the crosses show the following predicted ages: log($\tau/$yr) = 6, 7, 8, 9 and 10, 
starting from the upper-left.
The arrow shows the direction of reddening.
\label{fig-ub-vi}}
\end{figure}

Figure~\ref{fig-ub-vi} compares predictions from solar metallicity
stellar population models of Charlot \& Bruzual (2007;
hereafter CB07, private communication; see also \citet{BC2003}), 
with colors measured for our final cluster sample.  
Crosses mark the predicted colors for $10^6$, $10^7$, $10^8$, $10^9$, and $10^{10}$
yr clusters, starting from the upper left.  The arrow shows the
direction of reddening for a Galactic-type
extinction law \citep{Fitzpatrick1999}. 

The model predictions match
our cluster photometry relatively well.  The majority of the clusters
have blue integrated colors, indicating that they are fairly
young, with ages $\tau \lea\mbox{few}\times10^8$~yr.  At the top
of the diagram there are a few young clusters in crowded regions that fall above the
models, likely due to some contamination from neighbors in the lower
resolution $U$ band images (we see a similar effect for the colors of
clusters in M51; \citet{Chandar2010a}). The few sources with U-B colors significantly redder than the model
predictions are somewhat faint in the $U$ band, 
leading to large uncertainties in this color.  The bottom
panel of Figure~\ref{fig-ub-vi} shows the $B-V$ vs. $V-I$ two-color
diagram, and includes some objects for which no $U$ band photometry
was measured, either because they are outside the field of view or
because they are too faint in this band.  The sources which fall
significantly below (redward of) the model predictions in $B-V$ have
strong nebular line emission (which is observed in the
narrow-band F658N (H$\alpha$) filter), moving them off the model track.  While
two-color diagrams like those shown in Figure~\ref{fig-ub-vi} are useful
for visualizing the evolution of clusters, we actually use a $\chi^2$
minimization technique to estimate the ages of the
clusters.  This means that points outside the models (e.g., the
``high'' points in Figure~\ref{fig-ub-vi}) can be reasonably well fit.

We estimate the age $\tau$ and extinction
$A_V$ for each cluster as we have done in previous works
(see e.g., \citet{Fall2005} for details),
by performing a least $\chi^2$ fit comparing
observed magnitudes with the predictions from the CB07 stellar
population models with solar metallicity ($Z=0.02$), which
appear to match the measured colors of clusters in NGC~4449
reasonably well. The best fit combination of $\tau$
and $A_V$ for each cluster 
returns the minimum $\chi^2$: $\chi^2(\tau,A_V) =
\sum_{\lambda} W_{\lambda} (m_{\lambda}^{\mbox{obs}} -
m_{\lambda}^{\mbox{mod}})^2$, where the sum runs over all available
broad-band ($UBVI$) filters and the F658N narrow-band filter, but
requires a minimum of three measurements (including the $V$ band) to
estimate age and extinction. The weights $W$ are related to the
photometric uncertainty $\sigma_{\lambda}$ as $W_{\lambda} =
[\sigma_{\lambda}^2 + (0.05)^2]^{-1}$. The F658N filter includes both
stellar continuum and nebular line emission, and is dominated by line
emission from ionized gas
for the youngest ($\tau \lea \mbox{several}\times 10^6$~Myr)
objects, and by continuum emission from stars 
for clusters with ages of $\tau \gea
10^7$ yr. This enables us to use the narrow-band filter as a fifth
data point in many cases, regardless of the age of the cluster. 

We tested our method by estimating ages and masses from our $\chi^2$ analysis
for different assumptions regarding extinction and by
comparing our photometry with the predictions from a model with
lower metallicity ($Z=0.008$).
In general, we find that the age estimates are quite robust for
very young clusters $(\tau \lea 6\times10^6$~yr) that have nebular emission, 
and for clusters with ages between  $3\times10^7$~yr$\lea \tau \lea 10^9$~yr.
Clusters older than $\approx10^9$~yr suffer from the well-known age-metallicity degeneracy,
and we cannot tell if they are approximately a Gyr or older than $\approx10$ Gyr from
the data presented here. Eleven clusters in our catalog have 
integrated colors similar to those
of metal-poor globular clusters in our Galaxy (and in many other
galaxies) with $0.8 \leq V-I \leq 1.3$ and $0.6 \leq B-V \leq 0.9$.
The dating of clusters with no nebular emission and with ages
between $\approx6\times10^6$~yr and $\approx3\times10^7$~yr
is degenerate in some cases, with two nearly equally good
combinations of age and extinction.
We estimate that typical uncertainties are
$\approx0.3$ in log~$\tau$ for clusters with $\tau \lea 10^9$~yr,
typical for this method (e.g., \citet{Fall2009}).

The mass of each cluster is estimated from the total
$V$ band luminosity, corrected for extinction, and the
age-dependent mass-to-light ratios ($M/L_V$) predicted by the CB07 models.

\section{Formation and Disruption of the Clusters}

\begin{figure}[htp]
\epsscale{1.2}
\plotone{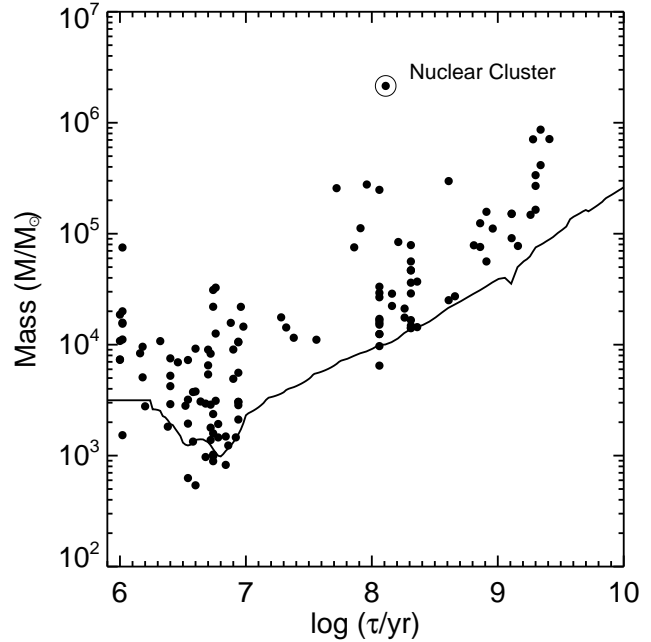}
\caption{Mass-age diagram of all clusters in our sample.
The solid line shows the approximate
 magnitude limit of $M_V=-7.0$.\label{fig-Mt}}
\end{figure}

The mass and age distributions of a population of star
clusters provides important clues to the formation and disruption of the clusters.
In Figure~\ref{fig-Mt} we show the mass and age estimates
for star clusters in NGC~4449. The solid line represents $M_V=-7$, the approximate
completeness limit of our sample, and shows that our sample
does not contain clusters over the same mass range at all ages,
because clusters fade over time.
The mass-age diagram shows some small scale features, such as 
a sparsely populated region
in the range $7.0 \lea \mbox{log}(\tau/\mbox{yr}) \lea 7.6$~yr.
This particular feature occurs where the predicted colors
loop back on themselves, covering a small region in color
space over a relatively long time, and effectively resulting
in a gap. This artificial, empty stripe and similar features
do not affect our conclusions.

Qualitative trends in the distribution of cluster ages 
and masses are apparent from the mass-age diagram.
We see that NGC~4449 has formed relatively
massive clusters ($M \gea 10^4~M_{\odot}$)
more or less continuously over the last $10^9$~yr.
A number of clusters (approximately half of our
sample), formed very recently, in the last
$\tau \lea 10^7$~yr, and a number of clusters clearly
formed $\approx 1-3\times10^8$~yr ago.
As mentioned in Section 3.3, our sample includes
$\approx 11$ candidate ancient globular clusters,
implying that NGC~4449 began forming stars
approximately a Hubble time ago.
This is broadly consistent with the results
of \citet{Annibali2008}, who found  that
stars in NGC~4449 have formed
over at least the last  $\approx10^9$~yr, 
with tentative evidence for earlier star formation as well,
based on the colors and luminosities of individual stars.
Qualitatively, the overall distribution of cluster ages
and masses in NGC~4449 appears to be similar to that in 
other galaxies with very different environments
and star formation histories, such as the merging Antennae
galaxies (e.g., \citet{Fall2005}, and the more
quiescent Magellanic Clouds \citep{Chandar2010a},
although the gap between $10^7-10^8$~yr is somewhat
more prominent.
We present a more quantitative analysis below.

\begin{figure*}
\epsscale{1}
\plottwo{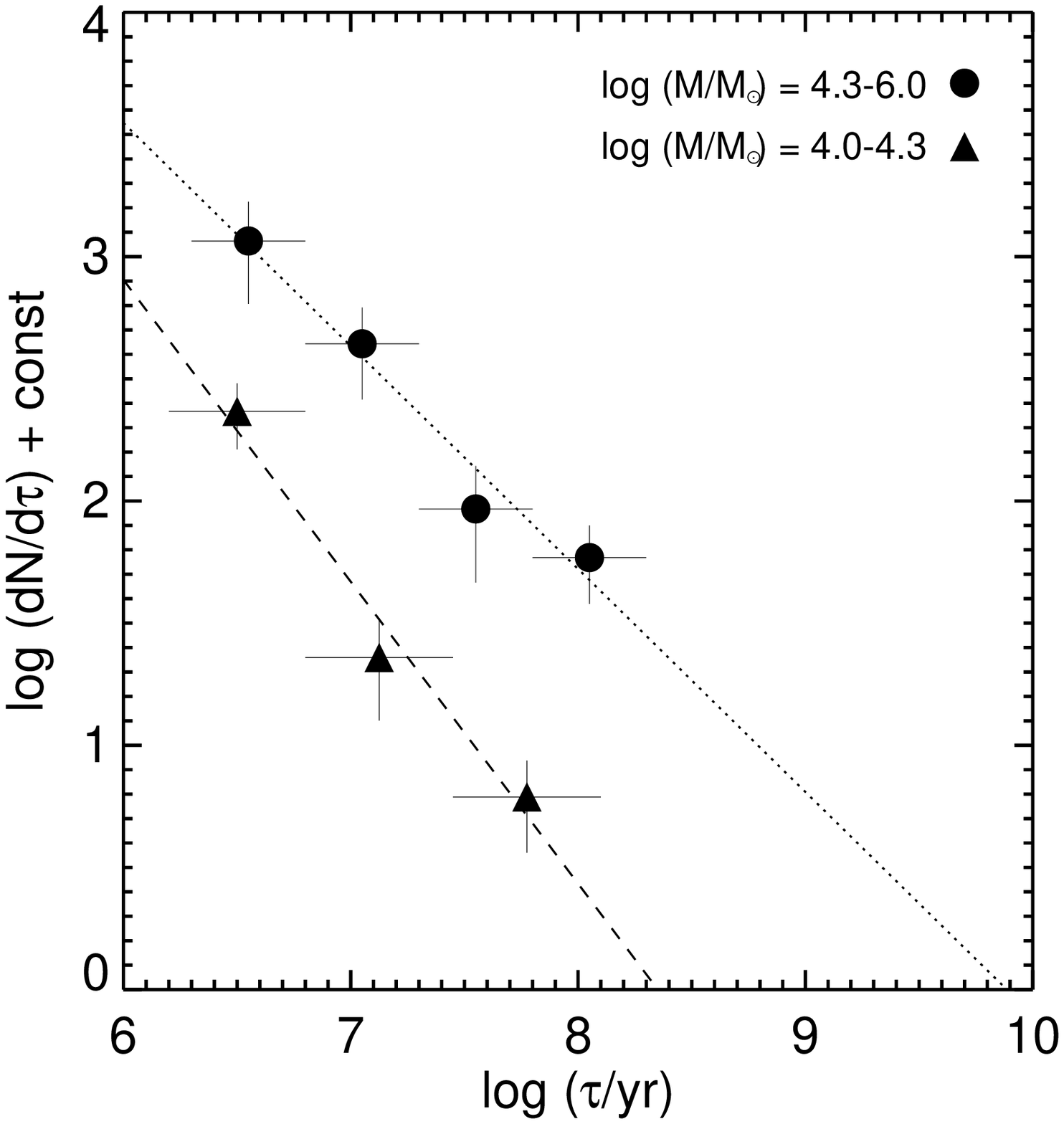}{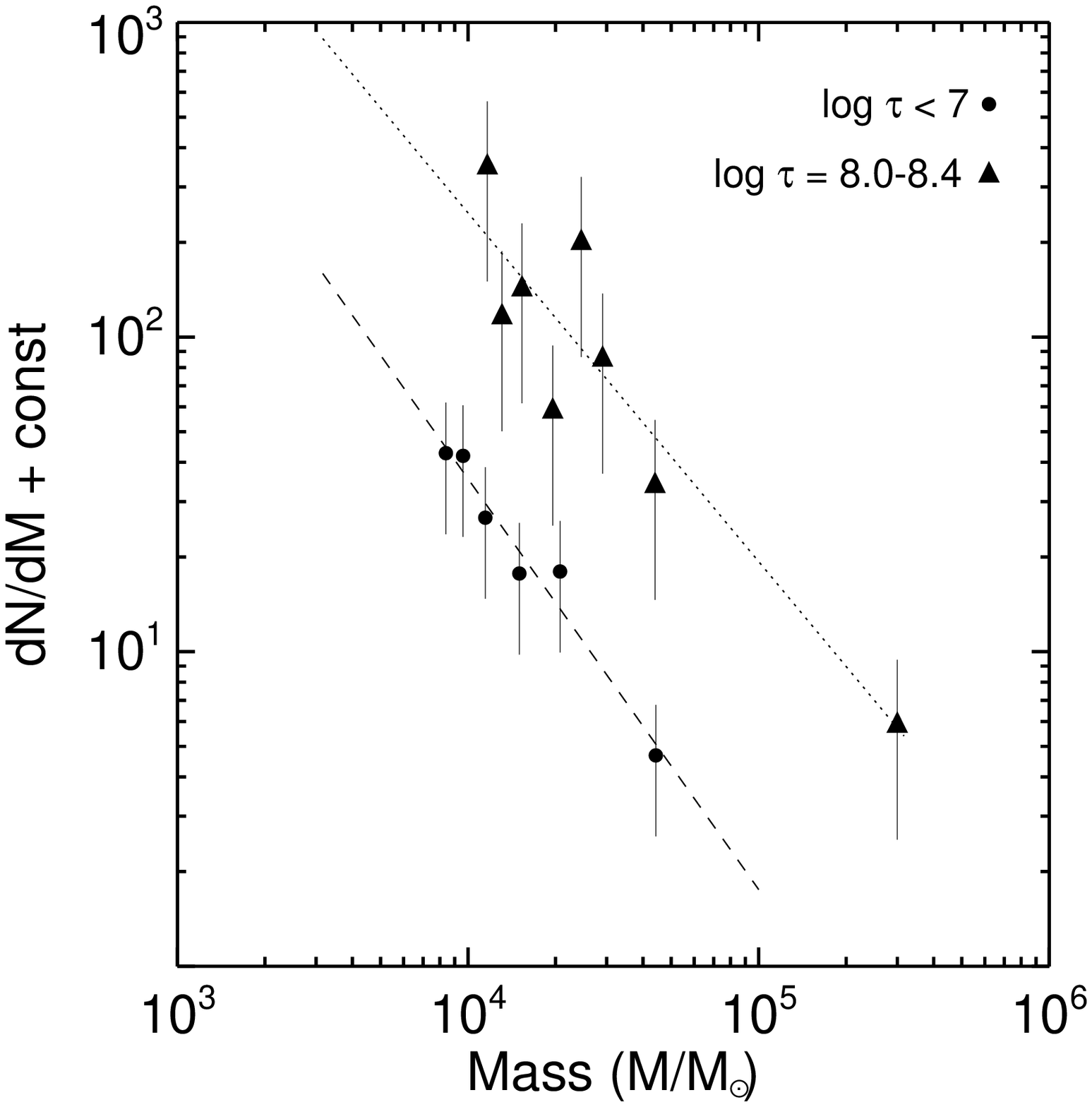}
\caption{The age distribution of clusters in NGC~4449 in
the indicated intervals of mass is shown on the left,
and the mass function in the indicated intervals of
age is shown on the right. 
The lines show the best fits to the distributions.
See text for more details.\label{fig-dndm}}
\end{figure*}

Figure \ref{fig-dndm}a shows the age distribution
for clusters in NGC~4449 in two different intervals of mass.
Note that the data points for each distribution are restricted to a mass-age range that is not affected by incompleteness (i.e. above the solid line), so the observed shapes are caused by the formation and disruption of the clusters and not by their fading out of our sample.
Although there are few data points, these distributions 
appear to decline steeply,
and can be approximated by a simple power law,
$dN/d\tau \propto \tau^{\gamma}$,
with best fits of $\gamma=-0.91\pm0.13$
for clusters more massive than $2\times10^4~M_{\odot}$,
and $\gamma=-1.21\pm0.21$ for clusters
with masses between $1-2\times10^4~M_{\odot}$.
These values of $\gamma$ are the same 
within the uncertainties.
We find that $\gamma$ can 
change by up to $\approx 0.3$ if different
bin sizes and centers are used,
but the overall decline starting at
very young ages does not change.
Based on these experiments, we take
$\gamma = -1.0 \pm0.3$.

Figure \ref{fig-dndm}b shows the mass function for clusters
in NGC~4449 in two different intervals of age.
These mass functions can be approximated by a power law,
$dN/dM \propto M^{\beta}$, with best fit
values of $\beta=-2.25\pm0.15$ for $\tau < 10^7$~yr,
and  $\beta=-2.07\pm0.19$ for $\tau = 1-2\times10^8$~yr,
which are the same within the uncertainties.
We take $\beta$ to be $-2.16 \pm 0.3$, the mean
and range of the values above.

Because the mass and age distributions appear to be
at least approximately independent of one
another, the joint distribution of cluster
ages and masses $g(M,\tau)$ can be approximated
as $g(M,\tau) \propto M^{\beta}\tau^{\gamma}$,
with $\beta \approx -2$ and $\gamma\approx -1$,
for $\tau \lea \mbox{few} \times10^8$~yr
and $M \gea 10^4~M_{\odot}$,
similar to the form observed for
clusters in the Magellanic Clouds \citep{Chandar2010a},
M83 \citep{Chandar2010b}, and in the Antennae
\citep{Whitmore2007,Fall2010}.

\citet{Chandar2010c} compiled results
for the age distribution of star clusters
in over a dozen different galaxies, including
dwarf irregular, spiral, and merging
galaxies.
All of these different galaxies appear to have similar
shapes to their (mass-limited) age distributions, where
they decline from the present to the past
(over the past $\approx$ few$\times10^8-10^9$~yr).
This shape for the age distribution
has previously been interpreted as due 
primarily to the gradual, early disruption rather than
to the formation of the clusters, since
it is far more likely that the clusters
in all of these galaxies have similar disruption
histories than it is that they have similar formation histories.

After their formation in the dense cores
of giant molecular clouds, different physical
processes cause clusters to lose mass and
to eventually disrupt. \citet{Fall2009} and \citet{Chandar2010a}
have suggested the following approximate sequence
and timescale for these processes: (1) removal of ISM by stellar feedback, $\tau \la 10^7$~yr;
(2) continued stellar mass loss, $10^7~ {\rm yr} \la \tau \la 10^8~ {\rm yr}$;
and (3) tidal disturbances by passing molecular clouds,
$\tau \ga 10^8$~yr. It is likely that the cluster mass and age, which can be approximated by power-law distributions, result from a complex situation that involves several of these disruption processes. We refer the interested reader to \citet{Fall2009} and \citet{Chandar2010a} for more details. On longer timescales, mass-loss is driven
by the escape of stars due to internal two-body relaxation, or evaporation.

\section{Spatial Correlation Between X-ray and Optical Sources}

\subsection{Correlation Between the Positions of XRBs and Star Clusters}

\citet {Kaaret2004} previously suggested that X-ray sources in three
different starburst galaxies (M82, NGC 1569 and NGC 5253) may have 
formed in young star clusters, because they are preferentially
located near star clusters, albeit with a significant displacement
($\approx 200$~pc on average). This conclusion was based on a
comparison between the cumulative distribution of displacements
measured between actual X-ray sources and their closest clusters,
and simulated displacements from a population of X-ray sources distributed
randomly relative to the clusters.
In order to explore whether or not X-ray binaries form in compact star clusters, 
here we compare the positions (of the entire sample) of 
23 candidate X-ray binaries (presented in Section 2)
with those of the star clusters (Section 3),
to look for a statistically significant spatial correlation between the two.

\begin{figure}
\epsscale{1.18}
\plotone{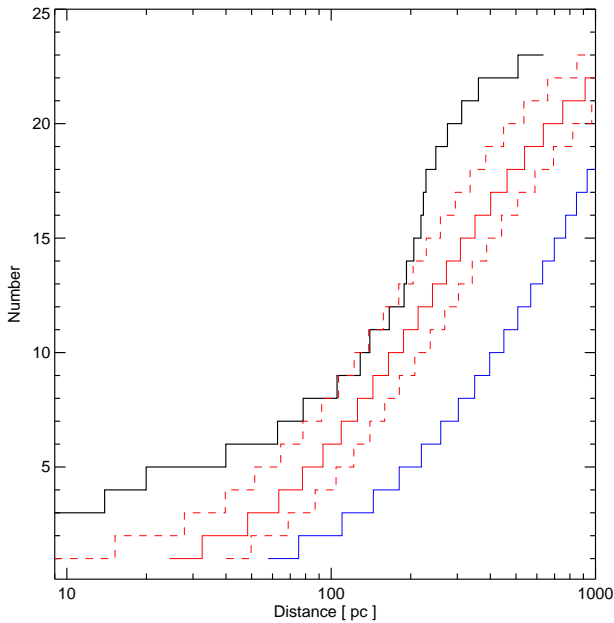}
\caption{Cumulative distribution of the displacement (in parsecs)
between X-ray binaries and star clusters.
The black (solid) line shows the result for sources in NGC~4449.
The other lines show results from our Monte Carlo simulations.
The red solid line is the median of 1000 random distributions where the
the locations of the X-ray sources are drawn from a distribution
that follows the light of the galaxy.
The dashed lines show the 1 $\sigma$ deviation from
this median curve. 
The blue solid line represents the median of 
1000 random distributions where the X-ray source locations
are drawn randomly from across the face of the galaxy.
\label{fig-X-ray-dist}}
\end{figure}

We repeat the spatial correlation analysis performed by \citet {Kaaret2004} 
for our sources in NGC~4449.
We identify the nearest star cluster to each X-ray binary,
and measure the (projected) distance between the two.
For XRBs that are not coincident with clusters, this gives a lower bound
on the displacement of the source from its parent cluster,
since the closest cluster is not necessarily the parent.
The results for NGC 4449 are shown in Figure~\ref{fig-X-ray-dist} as a cumulative distribution of
the separations between the binaries and the closest cluster.
The figure shows that there are 7 XRBs that are within
100~pc of a star cluster, and nearly all are
within $\approx$ 400~pc of a star cluster.
Interestingly, three HMXBs in NGC 4449 are spatially \emph{coincident}
with very young star clusters (within the $1\sigma$
positional uncertainty of $\approx 0.5\arcsec$ for $Chandra$), 
suggesting that these XRBs formed within
the clusters.

To determine whether the clustering of XRBs near
star clusters is statistically significant,
we compare the observations with simulated X-ray source
populations. We consider two different randomly distributed cases,
i.e. where the X-ray sources are {\em not} associated with the clusters.
First, we generate random sets of (23) uniformly distributed
sources, as done in \citet{Kaaret2004}, and find spatial displacements from the star
clusters using the procedure described above.
The simulation was run 1000 times, with the cumulative
distribution determined in exactly the same way as for the actual observations,
and the median result shown as the blue line in  Figure~\ref{fig-X-ray-dist}.
The uniformly distributed random sources show a clear
difference from the observed spatial displacements, similar to the results
found by \citet{Kaaret2004} for three different
starburst galaxies (not including NGC~4449).

Next, we select random sets of (23) sources distributed
in a more realistic fashion, one where the XRBs
fall off radially like the stellar light in NGC~4449
itself. This was accomplished by fitting the radial luminosity profile of
NGC 4449 in a 2MASS $K$ band image, by a broken power-law,
$\phi(L) \propto L^{\alpha}$, with $\alpha=-0.79$ over the radial range 150 $-$ 800 pc,
and $\alpha=-2.28$ over the radial range
800 $-$ 3000 pc. The $K$ band should give a reasonably good measure of
the distribution of older stars that form the
stellar backbone of NGC 4449, rather than
the distribution of young star clusters.
The median result from 1000 runs of 
the random simulations described above are shown as
the solid red line in Figure~\ref{fig-X-ray-dist}.
In this case, the results of the randomly drawn spatial
displacements are more similar to the observations, 
except at the small displacement end,
with the observations now overlapping partly with
the $1\sigma$ uncertainty line from the simulation
at separations $\gea 100$~pc. We find that a radial profile of
NGC~4449 determined from far-ultraviolet images taken with
GALEX gives very similar results to those
based on the K-band profile. Our results indicate that the specific recipe used to
populate synthetic X-ray sources in the image
can affect the interpretation of the results,
i.e. whether or not the X-ray sources are
associated with star clusters. In general, a uniform distribution is much
more likely to place synthetic X-ray sources in the
outer portions of the galaxy when compared with
the second method; because there are few X-ray sources
located in the outskirts of NGC 4449, this procedure
causes the simulation and observations to look
substantially different.

We conclude that, in the case where the sources are distributed
randomly but in a fashion similar to the underlying
galaxy light, the random distribution does not differ
significantly from the observed one,
{\em except} at the smallest separations,
where we find several XRBs closer to star
clusters than predicted by the random
distributions. These sources are discussed in more detail in Section~6.

\subsection{Optical Sources That are Coincident With HMXBs}

Three of the candidate HMXBs (X12, X21 and X46) have young  ($\tau \lea 10^7$ yr)
star clusters located within their $\approx0.5\arcsec$ positional uncertainties.
There is also an XRB coincident with an old
globular cluster (X29); this is almost certainly a low-mass X-ray binary (LMXB).
Here, we check the probability that this spatial
coincidence is due to chance superposition rather than to a physical
association between the HMXBs and the clusters. 
We distribute 23 XRBs (the number in our sample) randomly throughout NGC 4449 in
100,000 runs, and find one chance superposition approximately
every fifth run when all 129 clusters were included. 
This is $\approx$25 times less
than the four real observed coincidences between star clusters and XRBs. 
The probability of a chance superposition is 
almost identical if we only consider 
the three coincident HMXBs and star clusters younger than 10 Myr,
i.e. the most probable host clusters. 
Here, the random sources follow the
luminosity profile of NGC 4449 (see Section 4.1). If the random sources are
uniformly distributed instead, we find coincidences only $\sim$3\% of
the time, approximately 1 out of every 30 runs. Our results indicate
that the positional coincidence between the XRBs and clusters in these
three cases has a high statistical significance, and it is highly
unlikely the result of chance superposition.

In addition to the XRBs that are coincident with star clusters,
several others are coincident with an optical point source:
X6, X10, X14, X20, and X24, as presented in Section 3.2. 
These point sources are well within the body of NGC~4449.
They have the colors and magnitudes expected of supergiant stars, and
are almost certainly the donor star in the X-ray emitting binary.
Other candidate HMXBs are located in crowded regions, making it
more difficult to identify a unique optical counterpart to
the XRB, because there are several sources in the $HST$
image within the positional uncertainties of the X-ray source. 
Some of the X-ray sources (e.g., X18, X13, X20, X11, and X19)
do not have any obvious optical counterpart. 

\section{Discussion}

\subsection{Properties of Star Clusters Closest to HMXBs}

The ages of star clusters in the vicinity of the X-ray binaries can
help to constrain the ages and the nature of the HMXBs. Even when
there is no unambiguous stellar or cluster counterpart to an X-ray
source, it is plausible that the XRB is the same age as the stars and
clusters around it. Here, we present the physical properties of the
star cluster that is closest to each HMXB. In Table \ref{tbl-closest}, for each HMXB
we compile the distance to the closest star cluster in our catalog,
and the age and mass of this cluster.
We also include the X-ray luminosity of each source determined in
Section~2.1, and a column that lists whether the X-ray colors are better
described by a disk black body or a power law model, or are in the
absorbed portion of the diagram.

\begin{deluxetable*}{lccccccl}
\tablecolumns{7}
\tablecaption{Properties of Star Cluster Closest to XRBs in NGC~4449\label{tbl-closest}}
\tablewidth{0pt}
\tablehead{
\colhead{ID} & \colhead{X-ray} & \colhead{ID of Closest} & \colhead{Cluster Age\tablenotemark{1}} & \colhead{Distance} &  \colhead{Notes} \\
\colhead{} & \colhead{Model} & \colhead{Cluster} & \colhead{log($\tau$/yr)} & \colhead{(pc)}  & \colhead{ }
}
\startdata
X8 & Absorption & 96 & 9.34 & 266  & \\
X14 & Absorption & 117 & 8.06 & 339  & \\
X18 & ... & 117 & 8.06 & 285  & \\
X13 & Absorption & 112 & 8.06 & 232  & \\
X6 & Absorption & 93 & 6.64 & 122  & \\
X29 & ... & 89 & 8.91 & 7  & LMXB \\
X10 & Power-law & 99 & 8.31 & 286  & \\
X12 & Disk blackbody & 104 & 6.96 & 10  & Coincident \\
X20 & Absorption & 33 & 8.86 & 222  & \\
X11 & Absorption & 102 & 8.36 & 89  & \\
X23 & Disk blackbody & 66 & 6.58 & 181  & \\
X15 & Power law & 13 & 8.31 & 9  & Coincident \\
X24 & Disk blackbody & 69 & 6.84 & 68  & \\
X21 & Disk blackbody & 53 & 6.72 & 13  & Coincident \\
X28 & Disk blackbody & 81 & 6.72 & 136  & \\
X31 & ... & 87 & 6.70 & 630  & \\
X22 & Absorption & 70 & 6.38 & 133  & \\
X19 & Super soft source & 51 & 8.06 & 383  & \\
X41 & Super soft source & 50 & 6.90 & 225  & \\
X42 & Disk blackbody & 6 & 6.40 & 113  & \\
X43 & Absorption & 62 & 6.40 & 57  & \\
X44 & Absorbed & 75 & 6.16 & 134  & \\
X46 & Disk blackbody$-$Power law & 111 & 6.70 & 4  & Coincident \\
\enddata
\tablenotetext{1}{Uncertainties in the cluster age estimates are typically $\pm0.3$ in log~$\tau$.}
\end{deluxetable*}

The colors of HMXBs can vary, because these binaries go through
different emission states, or because of their geometric orientation,
since inclination effects through a disk can result in colors in the 
absorbed portion of the X-ray two-color diagram.
This means that it is not possible to definitively identify the
type of HMXB (e.g., black hole vs. neutron star) or to 
determine its age, just from the
X-ray properties. However, different types of HMXBs do show some
general trends in the color-color diagram.  For example, black hole
binaries often approximately follow the disk black-body models
\citep{Remillard_McClintock2006}. At high accretion rates the X-ray
emission from black hole binaries is dominated by thermal emission
from a disk, leading to both high luminosities and soft X-ray
colors. On the other hand, accreting low-mass neutron star binaries
often have absorbed power-law spectra.

\subsection{A Population of Very Young, Massive Black Hole Binaries in NGC~4449}

We can select very young binaries, which are likely to be high-mass black hole
binaries (BHB), in two different ways: (1) from their location
in the X-ray color-color diagram, and (2) from
their proximity to very young star clusters.
We first select candidate BHBs from Table \ref{tbl-closest}
based on their X-ray colors.
There are seven X-ray sources that are best
described by disk black-body models:
X12, X21, X23, X24, X28, X42 and X46.
All of these X-ray sources are either coincident with (X12, X21, and X46), 
or fairly close to (within 200~pc) a very young
$\tau \lea 8$~Myr star cluster.\footnote{Source X21 also happens 
to have an older, $\tau \approx 200$~Myr cluster nearby.}
The coincidence with clusters is particularly
important, because it establishes a direct connection
between the BHBs and very young clusters (recall
that in Section~5.1 we found that chance superpositions
are highly unlikely). These seven sources have a median X-ray luminosity of
$2.25\times10^{37}$ erg s$^{-1}$.
The fact that they have the X-ray colors expected for BHBs and 
are also close to very young star clusters strongly supports that these
sources are in fact, very young BHBs (discussed in more detail below).

Next, we select  candidate black hole binaries based solely on their proximity
(within 200~pc) of a very young ($\tau \lea 8$ Myr) star cluster.
These criteria return all seven XRBs
described by the disk black body models, plus 
four additional sources: X6, X22, X43, and X44.
All four of these are found in the ``absorbed'' portion of
the X-ray two-color diagram. However, they are located 
near star forming regions which contain very young $\tau \lea 8$ Myr
star clusters, and that are further out in NGC~4449 
(with galactocentric distances $R_{g} \gea 1.2$ kpc) on average
than the seven sources discussed previously.
The stellar density of the galaxy has dropped
significantly at these galactocentric distances, which
strongly suggests that X6, X22, X43, and X44 are associated
with the nearby, recent star formation.
In fact, random simulations return only a single XRB 
with $R_{g} \gea 1.2$ kpc and within 200 pc of a young star cluster in 10,000 runs.

The physical association between BHBs and very
young star clusters is also apparent as the small
separations in the cumulative distribution
of spatial displacements between XRBs and star
clusters (Figure \ref{fig-X-ray-dist}). This regime deviates strongly from random
distributions, as we would expect if 
BHBs are physically associated with very
young star clusters. We conclude that (at least some) BHBs form
in, and not just near, compact star clusters,
based on the fact that three candidate BHBs
in our sample are spatially coincident with
a very young star clusters and that a fourth is very
close to a young cluster (within a projected distance of 13~pc).

The ages of $\tau \lea 6-8$ Myr estimated for 
the coincident and proximate star clusters
to the 11 XRBs discussed above strongly suggests
that these are black hole (rather than neutron star) binaries. 
At these very young ages,
only stars initially more massive than $M \gea 25-30~M_{\sun}$ 
will have had time to become supernovae (based on
the Padova models for solar metallicity). 
While there is still some uncertainty about the
exact range of stellar masses that end their lives as black holes
rather than as neutron stars, most models predict that the transition
between these two types of compact remnants occurs for star with
initial masses somewhere in the range of $\approx 18-25~M_{\sun}$
(e.g., \citet{Fryer1999}), below the initial stellar masses that have
completed their main sequence lifetime in $\approx6-8$~Myr old star clusters.
However, metallicity effects may complicate the relationship between initial stellar mass and remnant type (e.g., \citet{Heger2003}). Given these uncertainties we will refer to these as 'candidate' BHBs.

One of the XRBs that is best described
by a disk blackbody model, X24, is not coincident with a star cluster,
but is coincident with a bright point source.
The lack of any other bright stars within the $1\sigma$ astrometric
uncertainty suggests that this is the high mass donor star in the XRB.
The luminosity and color of this source are consistent with
isochrones that are $\approx8-10$~Myr, as shown in Figure \ref{fig-cmd-stars},
similar to the age of the nearest star cluster.

To summarize, the main results of this section are that we find strong evidence
for a population of BHBs in NGC~4449, and that these massive
binaries likely formed recently in compact star clusters.
Three of the candidate BHBs appear to reside within their parent
star clusters, while the others do not. 
These very young ($\tau \lea 6-8$~Myr) sources
comprise a significant fraction of all
X-ray emitting binaries brighter than 
$\approx \mbox{few} \times 10^{36}$ erg~s$^{-1}$  in NGC~4449.
BHBs therefore make up approximately 48\% or 11 out of 23 XRBs.

\subsection{Processes Responsible for the Spatial Displacement Between BHBs and Star Clusters}

Black holes are the compact remnants of massive O stars.
O stars typically form in (massive) clusters rather
than in the field, although they can
be dynamically ejected from clusters into the field.
For example, studies of field O stars in the Milky Way \citep{deWit2005} show that the majority of field O stars appear to be runaways from nearby star clusters.
Moreover, many and possibly most O stars are born in binaries \citep{GM2001,Larson2001}.
Section~6.2 provided strong evidence that BHBs probably form in star clusters.

In the previous section, we found that while three of the candidate BHBs 
in NGC~4449 are coincident with a star cluster,  the majority of them are not.
What physical process(es) is responsible for this displacement
between BHBs and their parent clusters?
We consider three different mechanisms that could lead to the apparent
displacement between BHBs and their parent star clusters:

\begin{itemize}
\item The parent cluster has dissolved and is therefore
no longer visible, leading to an apparent displacement
between the BHB and neighboring clusters.

\item The BHB was ejected from its parent cluster during
dynamical interactions with stars in the dense
cluster core \citep{PRA1967,GB1986}.

\item  The BHB was ejected from its parent cluster due
to an asymmetric supernova kick \citep{Zwicky1957,Blaauw1961}.

\end{itemize}

In Section~4, we suggested that the shape of the cluster mass and age
distributions in NGC~4449 are primarily signatures of the disruption
rather than the formation of the clusters. Regardless of whether
individual clusters dissolve partially or fully, our results 
in NGC~4449 (and those in other galaxies), 
suggest that star clusters lose a significant
amount of mass very quickly, on timescales of only
$\approx10$~Myr.\footnote{This early mass loss
is not driven by relaxation of the cluster
due to two-body interactions (e.g., \citet{FZ2001}).  
Relaxation-driven evaporation operates
on significantly longer timescales than the ages of the BHBs, and ejects
mostly low mass stars.} UV spectra of starburst galaxies 
reveal that the dispersed field populations are dominated
by B stars, whereas UV-bright star clusters are often dominated by O
stars, consistent with a scenario of rapid cluster dispersal and/or
mass-loss (e.g., \citet{Tremonti2001,Chandar2003}). 
Despite growing evidence that even massive star clusters,
i.e., those most likely to host O stars,
may disperse rapidly, we believe that this is unlikely to be
the mechanism responsible for the observed spatial displacement between 
very young BHBs and star clusters.
N-body simulations show that an unbound cluster retains the appearance
of a bound cluster for $10-20$ crossing times \citep{BK2007}, 
on the order of 10-20 Myr for typical clusters in our NGC~4449 catalog.
Since no cluster is observed at the locations of the majority of the BHBs,
the early dispersal of clusters
can be ruled out as the origin of the
spatial displacement between very young, high mass BHBs and star clusters.

We conclude that many BHBs in NGC~4449 have been ejected from 
their parent clusters,
either via dynamical kicks due to interactions with other stars
in cluster cores, or due to an asymmetry in the supernova
explosion of the compact object, which nonetheless does not unbind the binary.
We cannot  differentiate between these two mechanisms
using the current observations.
We can however, estimate lower and upper limits to the ejection velocities.
We estimate a lower kick velocity by assuming that each BHB
was ejected from its closest cluster very soon after it formed,
and divide the distance to this cluster by
its age, neglecting for the moment uncertainties in
the cluster age estimates. These lead to lower limits of 
45, 10, 27, and 23 km/s for X23, X24, X28, and
X42, respectively. The other BHB binaries give limits
of 28, 55, 23, and 84 km/s for X6, X22, X43, and X44.
The cluster ages and hence the velocities are uncertain by
$\approx$ a factor of two, resulting in lower limits to the  velocities
between $\approx 5-160$~km/s.
Of course dynamical ejection need not have occurred 
right after the cluster formed, but could have occurred more recently.
We estimate upper limits to the kick velocities by assuming that
each BHB was ejected within the last 1~Myr, i.e. we 
divide the distance by 1~Myr.  This procedure gives a range
of velocities (for non-coincident sources) between $\approx 30-180$~km/s.

\subsection{The Nature of Older X-ray Binaries in NGC 4449}

In the previous subsections we found that $\approx 11$ of the 23
candidate XRBs in NGC~4449 are likely very young, massive BHBs.
The goal of this section is to better understand the
ages and types of the remaining X-ray binaries.

We first constrain the number of low mass
X-ray binaries (LMXBs) in NGC~4449, systems consisting 
of a black hole or neutron star accreting from a {\em low-mass}
companion. Source X29 is almost certainly
a LMXB, because it is spatially coincident with a cluster that has
integrated colors similar to those of ancient Galactic globular clusters. Its 
X-ray colors and luminosity however, are unremarkable when compared
with the rest of the XRB sample, and therefore cannot be used to
select ancient LMXBs in general. In their study of five early-type
galaxies (ellipticals and lenticulars), which have formed LMXBs but
not HMXBs, \citet{KMZ2007} find $\approx25-60$\% of all LMXBs are coincident 
with globular clusters and the rest are found in the field.
Assuming a similar fraction of cluster-to-field LMXBs in
NGC~4449 implies that there are $\approx 2-4$
field and hence  $\approx 3-5$ total LMXBs
in NGC~4449, $\approx 13 -22$\% of our total sample.

Our analysis implies that there are $7-9$ remaining X-ray binaries
in NGC 4449 that are neither very young, high mass BHBs nor very old
LMXBs. These are likely older
than $\approx10$~Myr and younger than several billion years (age of
LMXBs). Interestingly, from Table \ref{tbl-closest} we find that the
closest clusters to X14, X18, X13, X10, X20, X11, and X19 are all
intermediate age, with $\tau \approx100-400$~Myr. The separations
between these X-ray binaries and their closest clusters is larger than
those found between BHBs and their closest clusters, with a median
(mean) separation of $\approx$ 285~pc (250 pc).

These ``intermediate'' age HMXBs are likely neutron star X-ray
binaries with either a Be or supergiant star for a companion. Accretion
from the former occurs via a disk, while the latter are primarily wind
fed. There are also differences in the optical luminosities of Be
versus supergiant donor stars: 
Be-HMXBs in the Small Magellanic Cloud have $M_{V}$ ranging from $-2$
to $-5$ \citep{McBride2008}, which at the distance of NGC 4449
corresponds to $m_{V} \approx 26 - 23$, while supergiant HMXBs 
in the Galaxy are significantly brighter \citep{Chevalier_Ilovaisky1998}, with $M_{V}$
brighter than $\approx -6.5$  ($m_{V} \lea 21.5$). 
X10 and X14 are therefore candidate
supergiant-HMXBs, because they have bright coincident point sources (see Section~3.2).
X18, X13, X20, X11, and X19
however, do not show any bright coincident point sources in the optical
images down to $m_V  \approx 25$, and are therefore likely Be-HMXBs. 

Be-X-ray binaries dominate the population of HMXBs in the
Magellanic Clouds \citep{McBride2008}.  
They turn-on approximately 20-50 Myr after a star formation event, as the first
neutron stars are formed. The low mass of the neutron star and
lower accretion rates mean that Be-X-ray binaries typically have lower
X-ray luminosities than massive BHBs. However, most Be star
X-ray binaries go into outburst when the neutron star passes through
the circumstellar disk of the Be star and the accretion rate
increases. We note that the candidate Be-HMXBs in NGC 4449 have
X-ray luminosities considerably higher than quiescent Be X-ray
binaries. If they are indeed Be X-ray binaries,
then they are in outburst, and we are
detecting only the highest luminosity sources. This would suggest there is
a much larger population of Be X-ray binaries in NGC 4449 below our detection threshold.

In Section~6.2 we used the spatial displacement diagram to support
our conclusion that very young, massive BHBs form in compact star clusters.
Here, we note that there is an inherent limitation in our
ability to interpret this diagram for older binaries.
While the association between very young clusters 
and XRBs shows up clearly,
this is not true for intermediate age sources.
Figure 9 shows the cumulative distribution of
displacements between $\tau \gea 10$~Myr
X-ray binaries and star clusters,
where we have removed the BHBs and the known LMXB.
After $\approx 50-100$~Myr, XRBs that have been
ejected at even low velocities of a few km/s are no longer
obviously spatially associated with their parent clusters,
since at separations larger than $\approx 200$~pc
the distribution is consistent with the $1\sigma$
uncertainties from a random distribution.

\begin{figure}
\epsscale{1.18}
\plotone{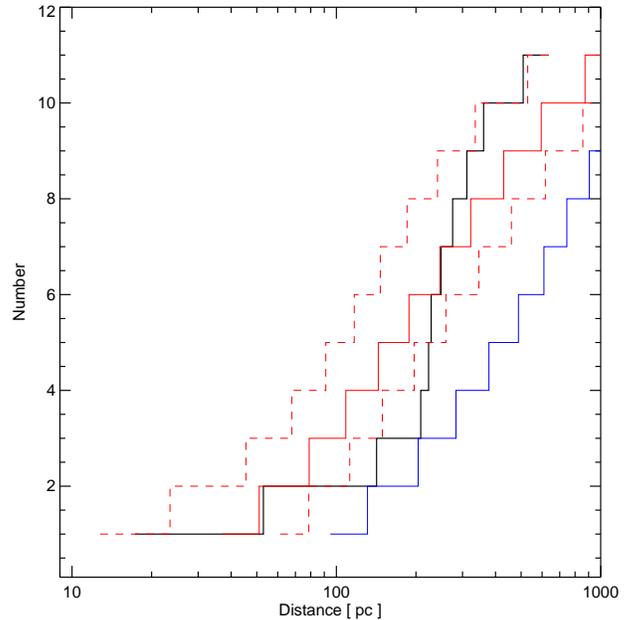}
\caption{Same as Figure~\ref{fig-X-ray-dist}, but excluding the eleven massive BHBs and the known LMXB.
\label{fig-9}}
\end{figure}

\section{Summary and Conclusions}

In this paper, we presented the discovery of 23
candidate X-ray binaries in the nearby starburst  galaxy NGC~4449,
 from \emph{Chandra}/ACIS-S observations.
We measured count rates, luminosities, and colors for these sources. 

We also presented a new catalog of 129 compact star
clusters brighter than $M_V \approx -7$
in NGC~4449 from multi-band, high resolution, optical imaging 
taken with the ACS/WFC and WFPC2 cameras on-board the $HST$.
This doubled the number of compact star clusters known in this galaxy.
Mass and age estimates for these clusters show that
NGC 4449 has formed relatively massive ($M \gea 10^4 M_{\odot}$)
clusters more or less continuously over the last $\approx10^9$ yr. 
The joint distribution of cluster ages and masses
appears to be similar to those found in other nearby
galaxies such as the Magellanic Clouds, M83, and the  merging Antennae galaxies,
albeit with somewhat larger uncertainties due to the smaller number of clusters,
and can be approximated as $g(M,\tau) \propto M^{\beta} \tau^{\gamma}$,
with $\beta= -2.16 \pm 0.30$ and $\gamma = -1.0 \pm0.3$.

Our main conclusion is that we have found clear evidence for a 
population of very young, high mass, black hole X-ray binaries in NGC~4449.
We find 11 candidate high mass BHBs, nearly half of the sample of X-ray emitting binaries.
Three of the BHB candidates are coincident, within the
astrometric uncertainties of $Chandra$, with a very young, $\tau \lea 8$~Myr star cluster,
and a fourth is nearly coincident with a very young cluster.
The others are all within 200~pc of very young clusters.
Based on these results, we suggest that these massive BHBs form
in star clusters, where most massive O stars (the progenitors of BHBs) are born.
Many are subsequently ejected from their parent clusters
either due to dynamical interactions within dense
clusters or as the result of an asymmetric supernova explosion.
The observed displacement between BHBs and very young
star clusters is not caused by the dissolution of the parent clusters.
The small separation between massive BHB candidates and their
closest cluster clearly deviates from
randomly populated distributions, further supporting our
conclusion that these X-ray binaries have a direct
relationship with young star clusters.

We found one X-ray binary in NGC~4449 that is coincident with an old
star cluster, and hence is almost certainly a LMXB.
Based on the fraction of LMXBs found in the field vs. in
globular clusters in early-type galaxies, we estimate
that there are $\approx 2-4$ field LMXBs and hence $3-5$
total LMXBs in NGC~4449, although they cannot be
identified from their X-ray properties alone.

We suggest that the remaining XRBs are intermediate age supergiant
and Be-HMXBs. Although none of these remaining binaries are coincident with
a star cluster, the closest cluster to these sources has an estimated
age between $100-400$~Myr.
In two cases, we identify coincident point sources in the $HST$ images
 that have luminosities consistent with those expected of 
supergiant donor stars.
The other candidate intermediate-age X-ray binaries do not have an
obvious optical counterpart, consistent with the expected
luminosities for Be star donors.

Based on these results, we conclude that 
high mass X-ray binaries, particularly massive, black-hole  binaries,
dominate the X-ray luminosity from NGC
4449. We suggest that the ages and locations of star clusters
provide important insight
and constraints on the different types of X-ray binaries
in nearby star-forming galaxies.

\acknowledgements 
We thank the anonymous referee, whose careful reading and helpful
suggestions significantly improved our manuscript.
This work was supported by NASA contract NAS8-39073 (CXC) and NASA AR8-9010.


\end{document}